\documentstyle[12pt,psfig,aasms4]{article}
\begin{document}

\title{Multi-wavelength study of the nebula associated with the galactic \\
 LBV candidate HD 168625 \altaffilmark{1}}

\author{A. Pasquali\altaffilmark{2},
A. Nota\altaffilmark{3,4}, 
L.J. Smith\altaffilmark{5},
S. Akiyama\altaffilmark{3}, 
M. Messineo\altaffilmark{6} \&
M. Clampin\altaffilmark{3}}

Submitted to {\it The Astronomical Journal}\\

\altaffiltext{1}{Based in part on observations made with the NASA/ESA Hubble Space
Telescope, obtained from the data archive at the Space Telescope Science
Institute. STScI is operated by the Association of Universities for
Research in Astronomy, Inc. under NASA contract NAS 5-26555.
Also based on observations with ISO, an ESA project with
 instruments funded by ESA Member States (especially the PI countries: France, 
 Germany), and with the participation of ISAS and NASA.
Also based on observations obtained at the Anglo-Australian Observatory, Siding Spring.}
\altaffiltext{2}{ESO/ST-ECF, Karl-Schwarzschild-Strasse 2, D-85748 Garching
bei M\"unchen, Germany; Anna.Pasquali@eso.org} 
\altaffiltext{3}{STScI, 3700 San Martin Drive, Baltimore, MD 21218, USA;
 nota@stsci.edu, clampin@stsci.edu}
\altaffiltext{4}{Affiliated with the  Astrophysics Division, Space Science Department of
ESA}
\altaffiltext{5}{Department of Physics and Astronomy, University College
London, Gower Street, London WC1E 6BT, UK; ljs@star.ucl.ac.uk}
\altaffiltext{6}{Leiden Observatory, P.O. Box 9513, 2300 RA Leiden, 
The Netherlands; messineo@strw.leidenuniv.nl}
 
\begin{abstract}
We present high resolution HST imaging of the nebula associated with the
galactic LBV candidate HD 168625, together with ISO imaging and AAT 
echelle spectroscopy. The overall nebular morphology is elliptical 
with the major axis at PA $\simeq$ 120$^{\deg}$. The dimensions of the nebula are
12$''$ x 16$''$.7 at H$_{\alpha}$ and 15$''$.5 x 23$''$.5 at 4 $\mu$m. 
In the HST H$\alpha$ image, the nebula is
resolved into a complex structure of filaments and arcs of different
brightness. The asymmetry is lost in the HST continuum image where
the nebula appears more diffuse and richer in filaments and clumps
with the shape of cometary tails. At 11.3 $\mu$m the nebular emission
peaks in two diametrically opposite lobes, placed on the nebula 
boundaries and along its major axis. 
A very faint loop is also visible at optical wavelengths, north and south of the shell. 
We suggest that the
nebula is an ellipsoid with projected sizes of 14$''$ and
9$''$ (0.19 pc $\times$ 0.12 pc) along the RA and DEC directions, 
respectively. This 
ellipsoid is expanding at 19 km s$^{-1}$ and is dynamically
as old as $\simeq$ 4800 yrs; it probably interacts with the
stellar wind and the loop so that PAH emission is detected from its caps,
i.e. the lobes seen in the ISO images. The chemistry of the
loop suggests that it is composed of un-processed material, probably
from the local interstellar medium swept by the stellar wind. 
\end{abstract}


\section{Introduction}
\label{sec:intro}
It is widely recognized that Luminous Blue Variables (LBVs) represent
a post-main sequence phase in which massive stars (M$_i \geq$ 20
M$_{\odot}$, Langer et al. 1994) lose
a considerable amount of mass via giant eruptions and minor outbursts.
The ejected gas and dust build up a circumstellar nebula chemically
enriched by the central star nucleosynthesis. From the expansion velocity
of known LBV nebulae, a dynamical age of a few 10$^4$ years is usually
inferred, which points to a very short-lived evolutionary phase. 
For this reason, LBVs are rare objects; indeed, only 40 are classified 
as such in the whole Local Group (Humphreys \& Davidson 1994). 
P Cygni and AG Carinae are considered the prototypes of the LBV class.
\par\noindent
The LBV nebulae provide us with a wealth of details about their
central stars. Their chemical composition is used to determine the 
evolutionary phase when the LBV instability triggers the nebula
ejection (cf. Smith et al. 1997 and Waters et al. 1999 in the
case of AG Carinae), and their morphology constrains the physics of
the central star wind.
Except P Cygni, all LBV nebulae display an asymmetric morphology, progressing
from elliptical (e.g. AG Carinae) to bipolar (e.g. Eta Carinae and HR
Carinae). Nota et al. (1995) reproduced the observed shapes through an
interacting wind model, where a spherical stellar wind interacts with a
pre-existing density contrast between the equatorial and polar direction. This
density contrast could be, for example,  induced by mass transfer in a binary
or rotation in a single star  (cf. Bjorkman \& Cassinelli 1993, Owocki et al.
1998). From the H$\alpha$ imaging it is also possible to estimate the ionized
gas  mass in the LBV nebulae which is a key parameter in the understanding of
the total mass ejected and of crucial importance  to  constrain  the evolutionary
models for massive stars. 
\par
Unfortunately, the diagnostic power of the nebular morphology is limited by
the spatial resolution accessible from the ground. Coronographic imaging has
so far been the observational technique achieving the highest resolution
possible from the ground: it has been able to resolve the global symmetry of
LBV nebulae out to the LMC  (Nota et al. 1995) and the nebular fine structure
only for close by objects,  such as AG Carinae. However, the  comparison with hydrodynamic
models (cf. Frank 1997, Garcia-Segura et al. 1997) obviously requires higher
levels of morphological details   to properly constrain the shaping mechanism
at work  in LBV nebulae. Further, a more complete analysis should rely on high
resolution imaging of ionized/neutral gas and dust; this would also assess the
total  (gas + dust) mass of LBV nebulae and hence the total mass lost by the
central star during the outburst. For these reasons, we have re-observed a
complete sample of LBV nebulae in the Galaxy and in the Large Magellanic Cloud
(Schulte-Ladbeck et al. 2002) with HST/WFPC2 and ISO/ISOCAM, among which is  
the galactic LBV candidate HD 168625. 
\par
HD 168625 is known to be variable with an amplitude of 0.06 mag (van
Genderen et al. 1992), although its variability does not closely follow
the typical pattern observed in the case of bona-fide LBVs. 
Nevertheless, its LBV candidacy was proposed when Hutsemekers et al. 
(1994), for the first time, resolved an associated  circumstellar nebula.
Nota et al. (1996) imaged the nebula with the STScI Coronograph
and resolved it into an  elliptical shell surrounded by two faint
filaments forming a northern and southern loop. A gas mass of $\sim$ 0.5
M$_{\odot}$ was derived from images in the light of H$\alpha$. 
Nota et al. also acquired two sets of spectroscopic data, six months apart, 
where HD 168625 was seen to fade by 0.3 mag and cool from T$_{eff}$ = 
15,000 K to 12,000 K at a constant mass-loss rate of $\sim$ 1.1 $\times$ 
10$^{-6}$ M$_{\odot}$yr$^{-1}$. 
In addition, the spectra were used to derive the
plasma properties of the nebula: for an assumed T$_e$ of 7,000 K, the average
density is 1000 cm$^{-3}$ and the nitrogen content is Log(N/H)$+$12 =
8.04 indicating that the nebula is composed of stellar ejecta. Unfortunately,
the spectra were taken at low resolution so that the kinematic structure
of the nebula was not fully resolved. Nota et al. could measure the
brighter, blueshifted edge of the nebula which apparently defined an
expansion motion of $\sim$ 40 km s$^{-1}$ and a dynamical age of
$\sim$ 10$^3$ yrs. 
\par\noindent
The nebula surrounding HD 168625 was also observed in the mid infrared
by Skinner (1997) and Robberto \& Herbst (1998). Their images (taken
between 4.7$\mu$m and 20$\mu$m) revealed emission by warm dust
in the eastern and western edges of the optical nebula. Skinner (1997)
derived a dust mass of 9.5 $\times$ 10$^{-5}$ M$_{\odot}$ for a
distance of 2.2 kpc. Robberto \& Herbst (1998) revised the distance
of HD 168625 from 2.2 to 1.2 kpc and estimated the dust mass of the
nebula to be $\sim$ 0.003 M$_{\odot}$. 
\par
We have ``revisited'' the nebula associated with HD 168625 employing
the high spatial resolution of WFPC2 onboard HST and the high
spectral resolution of UCLES on the AAT. These new data are complemented
with ISO/ISOCAM observations in order to derive a multi-wavelength,
detailed analysis of the nebular morphology and 
kinematics which are then used to trace back the outburst history
of HD 168625. 
The data are presented in Section~\ref{sec:data}. The HST and ISO images are
discussed in Sections~\ref{sec:hst} and ~\ref{sec:iso} respectively, and the
nebular  properties and kinematics are found in  Sections~\ref{sec:mass} and 
\ref{sec:neb}, and the stellar
spectrum from the echelle data-set is presented in Section~\ref{sec:star}. In
Section~\ref{sec:puzzle} we assemble and discuss the overall morphology of the 
HD 168625 nebula.

\section{Observations and data reduction}
\label{sec:data}
\subsection{The HST images}
HST/WFPC2 observed HD 168625 in 1997, on May 29 and  August 6.
The F547M and F656N filters were used 
for imaging in the V continuum (200 s) and the H$\alpha$ line (500 s)
to detect stellar emission reflected by dust and H$\alpha$ emission
from the nebular gas respectively.
The images were acquired in
CRSPLIT mode to remove cosmic rays and replace saturated pixels. 
Moreover, since they were taken four months apart, the May and August images 
have a relative rotation of 120$^\circ$ which allowed us to partially remove the
telescope spiders.
\par
The raw data were recalibrated  using the pipeline {\tt calwp2} in IRAF/STSDAS,
and the most appropriate reference files. No correction for the shutter
shading was applied since the adopted exposure times were longer than 
10 s. The five images from the same observation date, filter and orientation  
were combined with the {\tt crrej} task.
\par\noindent
The brightness of the central star saturated the five central columns of
each image. To remove them, we fit the nebula-free
portions of each image with a legendre function of order 1 along the X axis
using the IRAF task {\tt fit1d} and subtracted the fit from each
image.
\par
As the last step, the August images
were rotated and shifted to match the orientation of the May data and
then combined using the {\tt crrej} task.
We show in Figures 1 and 2 (Panel B) the final H$\alpha$ and V continuum 
images of the nebula associated with  HD 168625.
\par
During the reduction, we noticed the presence of two small regions of
saturated pixels.  One of them was at the center of HD 168625, and
the other 1.$''$15 from the image center. 
Assuming a distance to
HD 168625 of 2.8 kpc (cf. Sect. 3), this translates to 0.016 pc
or 3200 AU suggesting that HD 168625 might be a wide binary system.

\subsection{The ISO images}
HD 168625 was observed with ISOCAM, on board of ISO, on  September 9 and October
23, 1997. 
Images were taken on the star and on an adjacent region of sky with the filters 
SW5 and LW8. For these observations, we used a pixel size of 1.$''$5, which
yields an overall field of view of 48$''$ $\times$ 48$''$. The SW array is a InSb 32
$\times$ 32 pixels Charge Injection Device, while the LW detector is a  32
$\times$ 32 Gallium  doped Silicon photo-conductor array hybridized by Indium
bumps. The SW filter 5 covered the spectral region 3.1 -- 5.2 $\mu$m
where a number of Hydrogen emission lines from the nebular gas are
expected, while LW
filter 8 isolated the wavelength region 10.8 -- 11.9 $\mu$m centered on
PAH and dust emission. On source
integration times were 271 s for the SW5 observations  and  54 s  for the LW8
respectively. The gain was set to 2.0 for the SW5 images and to 1.0 for the
observation with the LW8 filter.

The images were recalibrated using the CAM Interactive Analysis Software (CIA).
For each filter, several steps were executed: first, we subtracted the
dark contribution.  An appropriate dark frame was selected from the archive,
consistent with the T$_{int}$ and gain of our observations. The images were then
{\it deglitched}, corrected for stabilization effects and divided by  an
appropriate flat
field. Finally they were converted into flux units and the sky was
subtracted. They are shown in Figure 3. 

\subsection{Ground-based spectroscopy}
High spectral resolution observations of the nebula surrounding HD 168625
were obtained with the UCL echelle spectrograph (UCLES)
and the MITLL2 $2048 \times 4096$ CCD detector at the coud\'e focus of the
3.9 m Anglo-Australian Telescope (AAT) on 20 and 21 September 1999
under 1.$''$0 seeing conditions. 
\par\noindent
A slit
of dimensions 1$''$ by 40$''$ was used with an interference filter
to isolate a single order covering H$\alpha$ and [NII]. The slit
was centred on the star and observations were obtained at position
angles of $0^\circ$ and $90^\circ$ (20 Sept) and $120^\circ$ (21 Sept)
with exposure times of 1800, 1500 and 1800s. We show in Figure 2
(Panel A)
the HST H$\alpha$  image of the nebula surrounding HD 168625 with superimposed
these three slit positions.
\par
The data were bias-subtracted, wavelength calibrated and sky-subtracted.
The spectral resolution is measured to be 6.9 km\,s$^{-1}$ 
and each spatial pixel corresponds to 0.$''$36.

\section{The distance to HD 168625}
\label{sec:dist}
The distance to HD 168625 is not well known. van Genderen et
al. (1992) assumed that the star is at the same distance as M17 or 2.2
kpc. This was questioned by Hutsem\' ekers et al. (1994) on the basis
that the nebular systemic velocity is greater than that of M17,
suggesting that HD 168625 may be more distant.  Robberto \& Herbst
(1998) derived a much closer
value of 1.2 kpc from infra-red photometric data and the model
atmosphere parameters obtained by Nota et al. (1996).  Since these
parameters were determined assuming a distance of 2.2 kpc, it
is not clear that the lower distance of 1.2 kpc is an independent
estimate. Moreover, if a distance of 1.2 kpc is assumed, the nebular
mass then becomes exceptionally low at 0.08 M$_\odot$ compared to
other LBV nebulae (Nota et al. 1995).

From our spectra we have measured a systemic LSR velocity for the star
of 25.5 km s$^{-1}$ (cf. Sect. 7) which indicates a kinematic distance of 2.8 kpc
assuming the Galactic rotation model of Brand \& Blitz
(1993). Hutsem\' ekers et al. (1994) noted that the Na I interstellar
lines in the spectrum of HD 168625 are at higher positive velocities
than those in the spectrum of the nearby bright star HD 168607,
assumed to be at the distance of M17. 
\par\noindent
Further evidence for HD 168625 not being at the same distance of M17
and HD 168607 comes from the proper motions measured by Hipparcos (cf. Table 1).
These clearly indicate opposite motion directions. Indeed, HD 168625 turns out to
move towards the NE while M17 and HD 168607 transit towards the SE. Therefore, we
agree with Hutsem\' ekers et al. (1994) that HD 168625 is not associated with either 
HD 168607 or M17 and probably is at larger distance than M17. Hereafter, we
adopt for HD 168625 a distance of 2.8 kpc which together with $E(B-V)=1.86$ (Nota et al. 1996)
and $V=8.4$ (van Genderen et al. 1992) gives $M_V=-9.6$.

\section{Optical nebular  morphology at high resolution}
\label{sec:hst}
The WFPC2 image of HD 168625 in H$\alpha$ light is shown in
Figure 1, where the stellar PSF has not been subtracted. As
already observed by Hutsem\' ekers et al. (1994) and subsequently Nota et al.
(1996) the bulk of the  nebular emission originates from an elliptical shell
with major axis oriented at PA $\simeq$ 120$^{o}$.  A larger, fainter loop is
visible in the northern region, as  already noted by Nota et al. (1996). 
A fragment of a southern loop is barely visible at PA $\simeq$ 250$^{o}$. In
the new HST images, the shell has a size of  size of $\sim$ 12$''$ $\times$
16.$''$7 (0.16 pc $\times$ 0.23 pc), 
in agreement with the previous measurements of  Nota et al. (1996).
A significant difference with the ground based coronographic images resides in
the  {\it texture}  of the shell, which is not uniform and homogeneous, but
rather a superposition of filaments  and arcs of different brightnesses, in a
way that is a close reminder the AG Carinae nebula (Nota et al. 1995). 
In the new
HST images we also resolve the structure of the brightest region in the
shell,  the southern rim, which appears fragmented, and resembles  the shape of
a letter {\it w}. In correspondence to  the vertex of the  letter {\it w} the
shell shows an inner and an outer layer, at different distances from the star
[3$''$ (0.04 pc) and 4.$''$3 (0.06 pc) respectively]. 

Diffuse emission is seen in the NE quadrant within the shell, between PAs
10$^{o}$ and 60$^{o}$: we will refer to this as the {\it paddle}. The {\it
paddle} is clumpy  and contains knots which are very similar in morphology to
the cometary-tail structures observed by HST/WFPC2 in the AG Carinae nebula
(Nota et al. 1995).
A second, prominent structure can be identified in the SE quadrant, which we
will label the {\it arm}: a filament  which  extends from the star towards
the edge of the shell and eventually merges  with an arc on the shell at 
$\simeq$ PA =  120$^{0}$. The arm is at $\simeq$ 30$^\circ$ from the shell major axis. 

\par\noindent The image of  HD 168625  taken in the V continuum light (F547M)
is shown in  Figure 2 Panel B. The difference between the H$\alpha$  and V
images is striking:  in the continuum image the shell  disappears  almost
completely, replaced  by a diffuse and filamentary nebula which has the same size
of the gaseous shell. Of the gaseous nebula, only the outline is barely
visible, but very different in appearance: the arcs and filaments have
disappeared to leave instead knots and clumps. Many more features similar to
the above mentioned cometary tails emerge in this image.  Neither loops are 
visible so that the {\it paddle} and the {\it arm} are left as the dominant
structures of the nebula in the V continuum light. 
To establish whether their higher V brightness (compared to H$\alpha$)
is intrinsic or due to different filter band-passes and exposure times,
we have performed aperture photometry on both the F656N and F547M 
images and for the southern edge of the shell, the {\it paddle} and the 
{\it arm}. The aperture radius was set to 5 pixel, for a full aperture
diameter of 0.$''$45. The aperture fluxes in units of counts have been 
summed overall the spatial extension of each feature, normalized by the 
exposure time of each filter (3000 s for the F656N and 800 s for the F547M 
filter) and multiplied by the filter inverse sensitivity in order to
transform them into units of ergs cm$^{-2}$ s$^{-1}$ A$^{-1}$. The
integrated, dereddened fluxes [E(B-V) = 1.7 from Nota et al. (1996)]  
are reported in Table 2 for the southern edge of the shell, the {\it arm} 
and the {\it paddle}.
\par\noindent
The dereddened flux in the F547M filter has been scaled by the ratio of the F656N to the
F547M bandwidth in order to compute the continuum flux
in the F656N filter under the assumption that the continuum is constant with wavelength.
This has been subtracted from the dereddened flux measured in the F656N and the
resulting emission in the H$\alpha$ line is reported in column 4 of Table 2.
Column 5 indicates the percentage contribution of the emission
in the H$\alpha$ line to the H$\alpha +$ continuum flux detected in the F656N filter.  
It can be seen that the H$\alpha$ flux is higher (81\%) in the shell southern edge and
decreases from the {\it arm} (64\%) and to the {\it paddle} (55\%).
The southern edge of the shell and the {\it paddle} are almost at 
the same distance ($\sim$ 4$''$, 0.05 pc) from the star so they should be
exposed to an equally diluted stellar radiation field. Their difference 
in the H$\alpha$ emission should then be due to either a different gas-to-dust
ratio or to a different gas density. Indeed, the H$\alpha$ flux is proportional
to the square of the gas density while the continuum flux is proportional to the
dust density which in turn can be expressed as the product of the gas density 
and the dust-to-gas ratio. Therefore, the H$\alpha$-to-continuum flux ratio
is proportional to the product of the gas density with the gas-to-dust ratio.
If the gas-to-dust ratio were constant across the nebula, the flux ratio would
depend on the gas density alone and hence the continuum flux. But this is
not the case for the shell southern edge and the paddle: from the former to
the latter the H$\alpha$-to-continuum flux ratio decreases and the continuum
flux increases. Therefore, we may qualitatively deduce that the gas-to-dust
ratio is not constant across HD 168625. 

\section{Mid-IR morphology and dust content}
\label{sec:iso}
Figure 3 shows the  final images of HD 168625
obtained with ISOCAM. The two images are taken with the SW5 and LW8 filters,
which cover  the spectral ranges 3.1 -- 5.2 $\mu$m and 10.8 -- 11.9 $\mu$m
respectively. In the images, North is up and East to the left.  The two images
present a very different view of the nebula. 
In the SW5 filter the central star
appears very bright while the nebula is faint. A faint diffuse, 
homogeneus emission is  discerned against the background, with an overall 
elliptical shape. The nebular emission originates from a region  $\simeq$
15.$''$5 $\times$  23.$''$5 (0.21 pc $\times$ 0.32 pc) 
in size, slightly larger than the optical shell. 
Allowing for the low resolution of the ISOCAM images compared with the
HST, the structure  observed well overlaps with the southern rim of the optical
shell. 

In the LW8 filter the star appears much fainter and the nebula much brighter,
with the same overall elliptical shape. In addition, two bright regions are easily
visible in the nebular emission, symmetrically located to the NW and to the SE
with respect to the central star. In the LW8 filter, the nebula has a 
larger extension, 31$''$ $\times$ 35.$''$5 (0.42 pc $\times$ 0.48 pc). 
The two brighter regions are located 
at the same distance from the central star of $\simeq$ 7.$''$5 (0.1 pc).
It is interesting to study the relative location of the features detected in
H$\alpha$ and in the mid-IR LW8 filter. In Figure 4 we show the 
composite final WFPC2 H$\alpha$ image with superimposed the  intensity contours
from the LW8 filter image. The two bright regions observed in the LW8 image
appear located adjacent to the  bright boundaries of the optical shell,
in correspondence to two {\it dark} regions. 
 
What is the origin of the mid-IR emission? \\
The spectral region of the SW5 filter (3.1 -- 5.2 $\mu$m) can trace PAHs (3.3
$\mu$m), continuum and  gas emission lines while the LW8 band (10.8 -- 11.9
$\mu$m) can indicate presence of SiC (11.4 $\mu$m) and  PAH (11.3 $\mu$m)
features (Skinner et al. 1997; Voors et al. 1997).

PAH  and silicate emission were first detected in the nebula around HD 168625 by
Skinner (1997).  Their published image of the HD 168625 nebula taken at 12.5 
$\mu$m displays a morphology which is identical to the one we observe in the LW8
filter, with two bright regions arranged symmetrically with respect to the star,
which the author interprets as a torus or  a disk. Skinner (1997) also shows
a spectrum of the central condensation  in the spectral range 8 - 24 $\mu$m
(their Figure 2). This spectrum shows PAH and silicate emission.

Robberto \& Herbst (1998) also observed the HD 168625 nebula at 4.7, 10.1, 11.6
and 19.9 $\mu$m. Two of these wavelenth regions, namely the 4.7 and the 11.6 $\mu$m are
directly comparable with the ISO SW5 and LW8 bandpasses. 
At 4.7 $\mu$m, Robberto \& Herbst barely detect the nebula, but well detect the
central star. The morphology of their 4.7 $\mu$m very faint nebular emission is very
similar, allowing for the different spatial resolution, to our SW5 image.
At 11.6 $\mu$m, Robberto \& Herbst (1998) find that the bulk of the emission is
coming from the nebula. They derive a size at 11.6 $\mu$m of 12$''$ $\times$
16$''$, which is consistent with our measurements if we consider the brightness
peaks. They also notice  that the emission is  concentrated in two bright arcs, in
the NW and SE quadrants, which correspond to the two bright regions we
see, at lower resolution, in the LW8 exposure. Robberto \& Herbst (1998) argue
that the arcs are the outer layer of a thin warm dust shell. A temperature of 
T$_{dust}= 135$ K provides the best fit of their measurements at 11.6, 19.9
combined with the IRAS points at 25 and 60 $\mu$m. 

It is interesting to notice that in their 11.6 $\mu$m image  (taken on August
28, 1996) the star appears very bright, while in our LW8 observation (taken 
in the fall of 1997) the star is barely detected.  Robberto \& Herbst (1998)
extract one-dimensional brightness profiles at various positions in the nebula
and clearly show that the peak intensity of the central star is only sligthly
lower than the peak intensity of the brightest regions in the nebula. 
Aperture photometry (with a 3$''$ radius) of the brightest regions in
the LW8 image gives an average flux a factor of
nearly 2 higher than elsewhere. Most likely, this discrepancy is due to the
already known variability of the central star (Nota et al. 1996). 

From our images, we  have measured an integrated nebular flux of 1.8 Jansky in
the SW5 filter  and 50.4 Jansky in the LW8 filter. The measurements have been
obtained by subtracting the contribution of the central star. These measured
values complement the data obtained by Robberto \& Herbst  and by IRAS and
have been used to estimate the temperature of the dust responsible for the
mid-IR emission. We have fit  to all data points black body curves at slightly different
temperatures, assuming a distance of 2.8 kpc and the nebular sizes as measured 
in the SW5 and LW8 images respectively.  A best fit black body
curve at 113 K is shown in 
Figure 5. If we give a lower weight to the IRAS 60 $\mu$m
measurement, which is assessed to be  ``uncertain or of lower quality'', the
data points are well fit by this black body curve, which is a few
degrees cooler, but not significantly different, from the best fit value of 
135 K derived by Robberto \& Herbst.

\section{Nebular Mass and Ionization Properties}
\label{sec:mass}
We have  used  the  integrated H$\alpha$ flux measured in  the new HST images
to reassess the determination  of the ionized gas  mass in the  nebula made by
Nota et al. (1996). The advantage of the high resolution provided by the HST
compared with the previous ground based coronographic images, is that the
stellar PSF is much smaller,  and, therefore, the  stellar contamination of
the surrounding nebular flux is also much smaller.   The integrated H$\alpha$
flux was measured in the new HST images by  masking  the residual halo from
the central star, and then integrating the H$\alpha$ flux in the image after
sky subtraction.  We have not corrected for the H$\alpha$ flux in the  masked
region of the image: consequently, the measured H$\alpha$ flux is a slight
underestimate of the true value. 

The reddened H$\alpha$ flux measured from the image is
4 $\times$10$^{-12}$~erg~s$^{-1}$~cm$^{-2}$. In order to deredden the nebular
measured flux, we have used our  value of E(B-V) = 1.86 $\pm$ 0.24 from Nota
et al. (1996). Dereddening the  measured H$\alpha$ flux, we obtain an
integrated flux of 9.6$\times$10$^{-10}$ erg s$^{-1}$ cm$^{-2}$ which is
slightly higher than the value of 3.7$\times$10$^{-10}$ erg s$^{-1}$
cm$^{-2}$  previously derived by Nota et al. (1996).  Adopting values of
T$_{e}$~=~7000 K, and n$_{e}$~=~1000 cm$^{-3}$ (Nota et al. 1996), we obtain
an ionized gas mass of 2.1 M$_{\odot}$. This  ionized gas mass is 
higher than the value of 0.5 derived by Nota et al. (1996) and considerably
higher than the mass of 0.04~M$_{\odot}$ calculated by Hutsemekers {\it et
al.} (1994).  However, a comparison with other LBV nebulae (Nota {\it et al.}
1995) shows the revised ionized gas mass of HD 168625 to be fairly typical.

Nota et al.(1996) concluded that the nebula was ionization bounded. In
fact, they estimated a Stromgren sphere radius of $\sim$ 0.1 pc.  If we
compare this value  with the physical size of the optical nebula obtained
adopting the new distance (0.16 $\times$ 0.23 pc) the nebula remains 
ionization bounded. It is interesting to notice that the spatial extent of
the {\it dusty} nebula, as imaged through the ISO/LW8 filter, is much larger
(0.42 $\times$ 0.48 pc), providing additional evidence that the optical nebula
is ionization bounded.

\section{Nebular kinematics}
\label{sec:neb}
We have extracted spatial profiles from the WFPC2 images of the HD 168625
nebula in correspondance with the slit positions that we used for 
spectroscopic observations. The tracings are presented in Figure 6:
the solid  lines refer to the images acquired in the H$\alpha +$
continuum while the dashed lines refer to the continuum. We have also associated 
the more prominent features with the shell southern edge, the {\it paddle} and the {\it arm}
described in Sect. 3.

The slits at PA = 90$^\circ$ and 120$^\circ$ intercept the shell on both sides
of the star; the slit at PA = 120$^\circ$ in particular detects both the
eastern and western lobes which dominate the IR emission of the nebula.
The slit at PA = 90$^\circ$ cuts through the western lobe and marginally
samples the northern faint end of the eastern one. The
slit at PA = 0$^\circ$ is centered on the southern edge of the shell where
the H$\alpha$ emission of the nebula peaks, but it also detects 
the northern loop and the fragment of the southern loop. 
\par
We have measured the radial velocity of the nebula with a spatial 
sampling of 2 pixels (0.$''$7) along the cross-dispersion
axis with a multi-gaussian fit of the nebular H$\alpha$ and 
[NII] $\lambda$6584 lines in each of the three spectra. 
Specifically, we have measured the peak wavelength and the
intensity of each component in the H$\alpha$
and [NII] $\lambda$6584 line profiles. Radial velocities have
been computed from the peak wavelengths and transformed into
heliocentric. We have typically detected 2 to 3 components
per position. 
\par
The overall distribution of the nebular [NII]6584/H$\alpha$ ratio
turns out to be bimodal, peaking at 0.51 $\pm$ 0.15
and 0.22 $\pm$ 0.05. The former is associated with the shell
and the latter with the northern and southern loop. 
\par\noindent
Figure 7 shows the shell velocity
distributions in Right Ascension and Declination where positions
are relative to the star. The top panel consists of all the
radial velocities measured in the spectra taken at PA = 90$^o$ and
120$^o$, and the bottom panel collects the radial velocities
determined in the spectra at PA = 0$^o$ and 120$^o$. 
The filled dots are the observed data
while the open circles represent the velocities obtained by 
fitting an expanding ellipse. The parameters of the best-fitting
ellipses are listed in Table 3 for both the velocity 
distributions in RA and DEC. The centre of both ellipses
is at (0$''$, 6 km s$^{-1}$). 
\par\noindent
According to the above parameters, the shell appears to be an
ellipsoid with a full linear size of 14$''$ (0.19 pc) and 9$''$ 
(0.12 pc) projected along the
the RA and DEC direction, respectively, and expanding at 19 km
s$^{-1}$. The spatial extensions derived by fitting the velocity
distributions well agree with those measured on the HST images.
Therefore, we estimate a lower limit to the dynamical age of $\simeq$ 4800 yrs
along the RA axis and $\simeq$ 3100 yrs in the DEC direction.
\par
The loop sample is restricted only to the Declination axis. Its
velocity distribution is less well sampled than the shell, and it can
be fit with either an expanding sphere or an expanding
ellipse. In Figure 8 we have plotted the observed data (filled
dots) and their best fits (open circles). In order to match the
loop sample, a circular shell should have a radius of about
9$''$.5 and expand at 17 km s$^{-1}$ with its centre at 
(0$''$, 4 km s$^{-1}$), while an ellipse should have a semi-minor
axis of $\simeq$ 9$''$ and expand at 19 km s$^{-1}$ (being its
centre at (0$''$, 6 km s$^{-1}$). The ellipse major axis (i.e.
the radial velocity axis) would also be tilted by $\simeq$ 27$^o$ 
towards South with respect to the shell ellipsoid. The loop
dynamical age would thus vary from $\simeq$ 7300 yrs as in the
case of a sphere to $\simeq$ 6200 yrs if the loop were
represented by a 2D ellipse.
\par
Because of the low spectral resolution of their data, Nota 
et al. (1996) could not distinguish between the shell and the
loop, and could only fit an expanding sphere to their 
velocity distribution, which in turn defined a radius of
about 12$''$ and an expansion velocity of $\simeq$ 37 km s$^{-1}$.

\section{A note on the stellar spectrum}
\label{sec:star}
We have extracted the stellar spectrum from the data at PA = 90$^\circ$,
normalized its continuum to unity and plotted it in Figure 9,
where the top panel shows the full flux-scale spectrum while
the bottom panel is a close up of the absorption lines. 

Two sets of lines can be identified in the stellar spectrum
of HD 168625 based on their FWHM: {\it i)} the H$\alpha$ P Cygni line and 
the CII
$\lambda\lambda$6578.1, 6582.9 \AA\/ and  FeI 6614 \AA\/ characterised
by a mean FWHM of 69 km s$^{-1}$; and {\it ii)} narrow absorption 
lines with an average FWHM of 10 km s$^{-1}$. The latter have been
identified with FeI, FeII, CaI, NiI features and possibly Mn and Ti
absorptions. These lines were also resolved in the spectrum of
P Cygni by Stahl et al. (1993). Israelian et al. (1996) detected 
DACs in the UV FeII and FeIII lines of P Cygni which vary over a
period of nearly six months and are believed to be associated with a
shell ejection. Therefore, the narrow absorption lines at optical
wavelengths may relate to the shell ejection mechanism as well.
In the case of HD 168625, it is interesting to compare the stellar
spectrum in Figure 8 with those obtained by Nota et al. (1996)
in 1995, five months apart. The discrete absorptions are present 
in the October spectrum but not in May; moreover, the star modified 
its H$\alpha$ profile from flat-topped to P Cygni and underwent
a variation of $\sim$ 0.3 mag in the continuum and $\sim$ 3000 K in 
T$_{eff}$ between May and October. This behaviour hints at a shell 
ejection; the fact that the narrow absorptions are also seen in the
1999 spectrum may indicate that shell ejections are a recurrent 
episode in the mass-loss history of HD 168625 as for P Cygni.
\par\noindent
We have used the CII $\lambda\lambda$6578.1, 6582.9 \AA\/ and  
FeI 6614 \AA\/ lines to derive the radial velocity of the star.
The stellar heliocentric radial velocity is 11.7 km s$^{-1}$
and the corresponding LSR value is of 25.5 km s$^{-1}$.

\section{Conclusions: Assembling the puzzle}
\label{sec:puzzle}
Multi-colour imaging of the nebula associated with HD 168625,
obtained from space with HST and ISO, shows an elliptical  nebula
 with the major axis oriented
at PA $\sim$ 120$^\circ$.
Its main structure is a shell which HST has resolved into nesting 
filaments. Both shell morphology and size depend on wavelength:
\par\noindent
{\it i)} in the H$\alpha$ light, the southern edge is well defined
while the northern part dissolves into diffuse circumstellar matter.
The shell surface brightness is higher along the southern edge and
peaks between PA $\sim$ 140$^\circ$ and PA $\sim$ 210$^\circ$.   
\par\noindent
{\it ii)} The shell  disappears in the V continuum light,
where it is replaced by diffuse, filamentary emission within a faint
outline which traces the boundaries of the gaseous nebula. At these 
wavelengths ($\lambda_c \sim$ 5480 \AA) the bulk of the emission seems 
to originate from a substructure, the {\it paddle},  to the NE of the central 
star.
\par\noindent
 {\it iii)} The shell lights up again at mid-infrared wavelengths, 3.1 -- 5.2
 $\mu$m and 10.8 -- 11.9 $\mu$m, but this time the emission peaks in two lobes
 which are adjacent to the inner edge of the shell, in correspondence to {\it
 dark or obscured} regions in the optical image. The {\it paddle} and the
 southern portion of the shell, which dominate the continuum and H$\alpha$
 emissions, are here far less pronounced. 
\par\noindent
 {\it iv)} The size of the nebula is
 different at optical and mid-IR wavelengths. In the light of H$\alpha$, the
 nebula has an extension of 12$''$ $\times$ 16.$''$7 (0.16 $\times$ 0.23
 pc), if we exclude the
 northern loop. At 3.1 -- 5.2 $\mu$m, the nebula has a larger extension of
 15.$''$5 $\times$  23.$''$5 (0.21 $\times$ 0.32 pc), and at 10.8 -- 11.9 
 $\mu$m, an even larger size of 31$''$ $\times$ 35.$''$5 (0.42 $\times$
 0.48 pc). 
\par
Aperture photometry of the optical images indicates that the
{\it paddle} and the shell southern edge may differ in the gas-to-dust content,
with the {\it paddle} more dusty and the southern edge of
the shell more gas-rich.
Aperture photometry of the infrared images suggests that the eastern
and western lobes of the shell are dominated by dust and PAH molecules,
in agreement with the findings of Skinner (1997) and Robberto \& Herbst
(1998) who resolved the lobes up to 20 $\mu$m.
\par\noindent
We believe that the nebular chemistry and kinematics derived here
can explain the observed composite morphology of HD 168625. 
\par\noindent
Nota et al. (1996) already derived for the nebula a Log(N/H) $+$
12 of $\simeq$ 8.04, showing that the nebula is indeed N-enriched
and hence of stellar origin. They could not resolve the nebula
into its components, as we did in this work with the help of echelle 
data. The higher spectral resolution of our data has allowed us to 
separate the shell from the loop and measure for each component
the [NII]6584/H$\alpha$ intensity ratio, that we use as a
N-abundance indicator. It turns out that the [NII]6584/H$\alpha$
ratio is, on average, 0.51 $\pm$ 0.15 for the shell and 0.22
$\pm$ 0.05 for the loop, which means that the shell is N-enriched
relative to the loop. In particular, the [NII]6584/H$\alpha$
ratio measured for the loop is very similar to what is observed
for galactic HII regions. We have indeed selected a number of
galactic HII regions from the sample of Shaver et al. (1983)
which are at the same distance of HD 168625 and/or have a plasma
temperature T$_e$([NII]) close to what assumed for HD 168625
(T$_e \sim$ 7000 K, Nota et al. 1996). We have computed their
relative [NII]6584/H$\alpha$ intensity ratios and determined
the mean value of 0.26 $\pm$ 0.08, which compares well to a
[NII]6584/H$\alpha \simeq$ 0.22 in the loop. 
This therefore implies that the loop is composed of un-processed
material which has been blown by the stellar wind. It is unlikely
that the loop is an interstellar bubble such the one detected
NW of HR Carinae (Nota et al. 1997): its dynamical age is at most 
$\simeq$ 7300 yrs old, too young for a HII region. The possibility
that the loop is a previous stellar ejection also seems improbable:
in this scenario, about 3000 years (i.e. the age difference between
the loop and the shell) would have been enough for the star to self-enrich
in N by almost a factor of 2. This time interval appears to be
quite short with respect to evolutionary models of B stars
(Lamers et al. 2001). Therefore, we suggest that the loop is local
interstellar medium swept by the stellar wind. 
The fact that we have been able to detect the loop 
only in the DEC direction may imply that the loop lies 
preferentially on a plane, maybe the equatorial plane of the
star. 
\par\noindent
The radial velocities measured for the shell indicate that it 
is an ellipsoid expanding at 19 km s$^{-1}$ in both
the RA and DEC directions. Its axes are 14$''$ (0.19 pc) 
and 9$''$ (0.12 pc) along the RA and DEC directions, respectively, 
in agreement with that estimated from the HST images. 
Such a morphology is very common among LBV nebulae, the best 
known example being AG Carinae.  
Nota et al. (1995) explained the elliptical shape of AG Carinae
by invoking a density contrast between the stellar equator and
poles which would restrict the nebula expansion on the stellar
equatorial plane and produce a ``waist'' in the circumstellar
nebula. The density contrast would leave the stellar polar axis as 
the only free direction to the nebular expansion and
therefore would shape the nebula into an ellipsoid. There exist
several ways to produce a density contrast (Livio 1995), such 
as stellar rotation and stellar binarity. Although no solid 
evidence for stellar rotation or binarity is available for
HD 168625, one of these mechanisms could be 
responsible for the overall morphology of the nebula associated
with HD 168625 and also would preferentially direct
the present stellar wind along the polar axis of the star, i.e.
the RA axis of the nebular ellipsoid, so that it would interact
with the ``caps'' of the ellipsoid. The interaction would destroy
the CO molecules and give rise to PAH emission as 
detected by ISO in the lobes of the nebula. Moreover, the HST
images suggest that the shell caps may be interacting with the loop 
and this could also produce PAH emission. 

\acknowledgements
We wish to thank the referee for his/her valuable comments
and suggestions.

\newpage
\begin{deluxetable}{lccl}
\tablecaption{Proper motions of HD 168607, HD 168625 and M17.}
\tablewidth{0pt}
\tablehead{
\colhead{} & \colhead{$\mu_{\alpha}$ (mas/yr)} & \colhead{$\mu_{\delta}$
(mas/yr)}
& \colhead{References}
}
\startdata
HD 168607 & 1.18 $\pm$ 1.24 & -1.16 $\pm$ 0.81 & Perryman et al. (1997)\nl
HD 168625 & 0.95 $\pm$ 1.23 & +0.04 $\pm$ 0.84 & Perryman et al. (1997)\nl
M17       & 0.14 $\pm$ 1.49 & -3.63 $\pm$ 0.94 & Baumgardt et al. (2000)\nl
\enddata
\end{deluxetable}

\begin{deluxetable}{lccccc}
\tablecaption{Dereddened F656N and F547M photometry of the southern edge of the
shell,
the {\it arm} and the {\it paddle}. An E(B-V) = 1.7 is adopted from Nota et al.
(1996).
}
\tablewidth{0pt}
\tablehead{
\colhead{} & \colhead{F656N} & \colhead{F547M} & \colhead{Continuum} &
\colhead{F656N -
 F547M} &
\colhead{H$\alpha$/(H$\alpha +$ cont.)}\nl
\colhead{} & \colhead{H$\alpha +$ cont.} & \colhead{cont.} &
\colhead{in F656N}
 & \colhead{}
& \colhead{}
}
\startdata
Southern edge & 1.6 $\times$ 10$^{-16}$ & 1.1 $\times$ 10$^{-16}$ & 2.9
$\times$ 10$^{-
17}$
& 1.3 $\times$ 10$^{-16}$ & 81\% \nl
Arm           & 6.8 $\times$ 10$^{-18}$ & 9.1 $\times$ 10$^{-18}$ & 2.4
$\times$ 10$^{-
18}$
& 4.4 $\times$ 10$^{-18}$ & 64\% \nl
Paddle        & 8.2 $\times$ 10$^{-17}$ & 1.4 $\times$ 10$^{-16}$ & 3.7
$\times$ 10$^{-
17}$
& 4.5 $\times$ 10$^{-17}$ & 55\% \nl
\enddata
\end{deluxetable}

\begin{deluxetable}{lccccc}
\tablecaption{Parameters of the best-fitting expanding ellipses.}
\tablewidth{0pt}
\tablehead{
\colhead{} & \colhead{Semi-major Axis} & \colhead{Semi-minor Axis}
}
\startdata
Distribution in RA & 19 km s$^{-1}$ & 7$''$\nl
(PA = 90$^o$, 120$^o$) &            &\nl
Distribution in DEC & 19 km s$^{-1}$ & 4.5$''$\nl
(PA = 0$^o$, 120$^o$) &             &\nl
\enddata
\end{deluxetable}

\clearpage
\begin{figure}
\caption[]{The H$\alpha$ image of HD 168625 taken with WFPC2 in the
F656N filter. North is up and East on the left. Panel B: the same as in
panel A with superimposed the slit positions for the spectroscopic
observations. Panel C: HD 168625 in the V continuum light, as observed
with WFPC2 in the F547M filter. North is up and East on the left.}
\end{figure}
\clearpage
\begin{figure}
\caption[]{Panel A: The H$\alpha$ image of HD 168625 taken with WFPC2 
with superimposed the slit positions for the spectroscopic
observations. Panel B:  HD 168625 in the V continuum light, as observed
with WFPC2 in the F547M filter. North is up and East on the left.}
\end{figure}

\clearpage
\begin{figure}
\caption[]{Upper panel: The nebula associated with HD 168625 imaged with
ISO through the SW5 filter ($\lambda_c$ = 4.0 $\mu$m). North is
up and East on the left. Bottom panel: HD 168625 as it appears at 11.3
$\mu$m with the LW8 filter of ISO. North is up and East on the left.}
\end{figure}
\newpage
\begin{figure}
\caption[]{Composite final WFPC2 H$\alpha$ image with superimposed the  intensity
contours from the LW8 filter image. North is up and East on the left.}
\end{figure}
\begin{figure}
\caption[]{Black body curve  (T = 113.5 K) which best fits  the IR
data points from this paper, Robberto \& Herbst (1999) and IRAS (bands at 
25$\mu$m and 60 $\mu$m). }
\end{figure}
\begin{figure}
\caption[]{Spatial profiles of the nebula from the H$\alpha$ (solid line)
and V continuum (dashed line) images as a function of distance from
the star in arcsec. The more prominent features are labelled according
to Figure 1.}
\end{figure}
\begin{figure}
\caption[]{Distribution of the shell expansion velocities
as a function of distance from the star (in arcsec) in
Right Ascension and Declination. Filled dots
represent the measured velocities while open dots are computed from the
model of an expanding ellipse.}
\end{figure}
\begin{figure}
\caption[]{Distribution of the loop expansion velocities
as a function of distance from the star (in arcsec) in
Declination. Filled dots
represent the measured velocities while open dots are computed from the
model of an expanding sphere (top panel) and an expanding ellipse
(bottom panel).}
\end{figure}
\begin{figure}
\caption[]{The stellar spectrum of HD 168625 extracted from the slit
at PA = 90$^\circ$. The full-flux scale spectrum is plotted in the
top panel where the broad emission/absorption lines are identified.
A close-up of the same spectrum is shown in the bottom panel where
the narrow absorption lines are labelled.}
\end{figure}

\end{document}